\pdfoutput=1

\documentclass[10pt,nofootinbib,twocolumn,secnumarabic,amssymb,aps,groupedaddress,prl,preprintnumbers,longbibliography]{revtex4-1}

\ifnum\pdfoutput=1\else
\PassOptionsToPackage{hypertex}{hyperref}
\fi
\usepackage{hyperref}

\def\BernFiniteness{Bern:2007hh,*Bern:2008pv,*Bern:2009kf,*Bern:2009kd,*Dixon:2010gz}
\def\StringyCounterterms{Berkovits:2006vc,*Green:2006gt,*Green:2006yu,*Green:2008bf,*Berkovits:2009aw}
\def\superspace{Howe:2002ui,*Stelle:2007zz,*Bossard:2009sy,*Stelle:2009zz}
\def\Unitarity{Bern:1994zx,*Bern:1994cg,*Bern:2004cz}
\def\SUBnonanomalous{Marcus:1985yy,*diVecchia:1984jh}
\def\Eisenstein{Green:2010wi,*Green:2010sp,*Green:2010kv}
\def\StSt{Stieberger:2009rr,*Stieberger:2007am,*Stieberger:2007jv}
\def\grosssloan{Gross:1986iv,*Gross:1986mw}
\def\NF{Bianchi:2003wx,*Beisert:2003te}
\def\shorts{Flato:1983te,*Dobrev:1985qv,*Binegar:1985ib,*Morel:1985if}
\def\LBCounterterms{Howe:1980th,*Kallosh:1980fi}
\def\ecite#1{\edef\ecitetmp{#1}\cite{\ecitetmp}}

\ifx\hypersetup\sadfkjashdfkxja\newcommand\hypersetup[1]{}\newcommand{\texorpdfstring}[2]{#1}\fi
\hypersetup{pdftitle={E7(7) constraints on counterterms in N=8 supergravity}}%
\hypersetup{pdfkeywords={}}%
\hypersetup{pdfsubject={}}%
\hypersetup{pdfauthor={N. Beisert, H. Elvang, D. Freedman, M. Kiermaier, A. Morales, S. Stieberger}}
\hypersetup{plainpages=false}
\hypersetup{pdfpagemode=UseNone}
\hypersetup{bookmarksnumbered=true}
\hypersetup{pdfstartview=FitH}
\hypersetup{colorlinks=false}
\hypersetup{citebordercolor={.5 1 .5}}
\hypersetup{urlbordercolor={.5 1 1}}
\hypersetup{linkbordercolor={1 .7 .7}}

\ifx\href\asklfhas\newcommand{\href}[2]{#2}\fi

\usepackage{amssymb,amsbsy,amsfonts,amssymb,amsmath}

\let\oldbfseries=\bfseries
\let\oldmdseries=\mdseries
\let\oldnormalfont=\normalfont
\renewcommand{\bfseries}{\oldbfseries\boldmath}
\renewcommand{\mdseries}{\oldmdseries\unboldmath}
\renewcommand{\normalfont}{\oldnormalfont\unboldmath}

\newcommand{\cn}{{\cal N}}
\def\<{\langle} 
\def\>{\rangle} 
\def\hp{\mathord{+}}
\def\hm{\mathord{-}}
\newcommand{\dash}{\text{-}}

\newcommand{\smallsection}[1]{\vspace{-2ex}\subsection{#1}\vspace{-2ex}}

\usepackage{color}
\usepackage{cancel}

\definecolor{darkred}{rgb}{0.5,0,0}
\definecolor{darkblue}{rgb}{0,0,0.4}
\definecolor{darkgray}{rgb}{0.3,0.3,0.3}
\definecolor{darkgreen}{rgb}{0.,0.23,0.1}
\def \bdgreen{\color{darkgreen}}%
\def\bdr{\color{darkred}}%
\def\bdb{\color{darkblue}}%
\def\blb{\color{darkblue}}%

\def\be{\begin{equation}}
\def\ee{\end{equation}}

\begin{document}

\preprint{AEI-2010-146}
\preprint{MCTP-10-40}
\preprint{MIT-CTP-4176}
\preprint{MPP-2010-122}
\preprint{PUPT-2348}

\title{\texorpdfstring{$E_{7(7)}$}{E7(7)} constraints on counterterms in \texorpdfstring{$\cn=8$}{N=8} supergravity}

\author{N. Beisert{}$^a$}
\author{H. Elvang{}$^{bc}$}
\affiliation{}
\author{D. Freedman{}$^{de}$}
\affiliation{}
\author{M. Kiermaier{}$^f$}
\affiliation{}
\author{A. Morales{}$^d$}
\affiliation{}
\author{S. Stieberger{}$^g$}
\affiliation{}
\vspace{2mm}
\author{}
\affiliation{{}$^a$Max-Planck-Institut f\"ur Gravitationsphysik,
Albert-Einstein-Institut,
Am M\"uhlenberg 1, 14476 Potsdam, Germany}

\affiliation{{}$^b$Michigan Center for Theoretical Physics (MCTP), \\ Department of Physics, University of Michigan, Ann Arbor, MI 48109, USA}

\affiliation{${}^c$Institute for Advanced Study, Princeton, NJ 08540, USA}

\affiliation{${}^d$Department of Mathematics and ${}^e$Center for Theoretical Physics, Massachusetts Institute of Technology, Cambridge, MA 02139, USA}

\affiliation{${}^f$Department of Physics, Princeton University, Princeton, NJ 08544, USA\,,}

\affiliation{${}^g$Max-Planck-Institut f\"ur Physik, Werner-Heisenberg-Institut, 80805 M\"unchen, Germany}

\date{\today}

\begin{abstract}
We prove by explicit computation that 6-point matrix elements of $D^4R^4$ and $D^6R^4$ in $\cn=8$ supergravity have non-vanishing single-soft scalar limits, and therefore these operators violate the continuous $E_{7(7)}$ symmetry.
The soft limits precisely match automorphism constraints. Together with previous results for $R^4$, this provides a direct proof that no $E_{7(7)}$-invariant candidate counterterm exists below 7-loop order.
At $7$-loops, we characterize the infinite tower of independent supersymmetric operators $D^4\!R^6$, $R^8$, $\varphi^2 R^8$,$\ldots$ with $n>4$ fields and prove that they all violate $E_{7(7)}$ symmetry.
This means that the 4-graviton amplitude determines whether or not the theory is finite at 7-loop order.
We show that the corresponding candidate counterterm $D^8R^4$ has a non-linear supersymmetrization such that its single- and double-soft scalar limits are compatible with $E_{7(7)}$ up to and including 6-points.
At loop orders $7, 8, 9$ we provide an exhaustive account of all independent candidate counterterms with up to $16, 14, 12$ fields, respectively, together with their potential single-soft scalar limits.
\end{abstract}


\maketitle

\section{Introduction}
\vspace{-1mm}
$\cn=8$ supergravity has maximal supersymmetry, and the classical theory has global continuous $E_{7(7)}$ symmetry which is spontaneously broken to $SU(8)$.
Explicit calculations have demonstrated that the 4-graviton amplitude in $\cn=8$ supergravity is finite up to $4$-loop order \ecite{\BernFiniteness}. Together with string- and superspace-based observations \ecite{\StringyCounterterms,\superspace}, this spurred a wave of renewed interest in the question of whether the loop computations based on generalized unitarity~\ecite{\Unitarity} could yield a UV finite result to all orders\footnote{This question is well-defined whether or not $\cn=8$ supergravity is sensible as a full quantum theory \cite{TomTalk,GOS}.} --- or at which loop order the first divergence might occur.

In gravity, logarithmic UV divergences in on-shell $L$-loop amplitudes are associated with local counterterm operators of mass dimension $\delta=2L+2$ composed of fields from the classical theory. The counterterms must respect the non-anomalous symmetries of the theory. It was shown in \cite{Harmonic,EFK,NonHarmonic} that below 7-loop order, there are only 3 independent operators consistent with linearized $\cn=8$ supersymmetry and global $SU(8)$ R-symmetry \ecite{\SUBnonanomalous}. These are the 3-, 5- and 6-loop  supersymmetric candidate counterterms $R^4,  D^4R^4$, and $D^6R^4$.

The perturbative $S$-matrix of $\cn=8$ supergravity should respect $E_{7(7)}$ symmetry \cite{E77nonanomalous}, so one must  subject $R^4,  D^4R^4$, and $D^6R^4$ to this test. A necessary condition for a counterterm to be $E_{7(7)}$-compatible, is that its matrix elements vanish in the `single-soft limit' $p^\mu \to 0$ for each external  scalar line \cite{BEF,Nima,KalloshKugo}. The scalars of $\cn=8$ supergravity are the `pions' of this soft-pion theorem since they are the 70 Goldstone bosons of the spontaneously broken generators of $E_{7(7)}$. It was recently proven \cite{EK}  that the soft scalar property  fails for 6-point matrix elements of the operator $R^4$ (see also~\cite{BD}). Thus $E_{7(7)}$ excludes   $R^4$ and explains the finite 3-loop result found in \cite{\BernFiniteness}.

In the present paper we show first that the 5- and 6-loop operators  $D^4R^4$ and $D^6R^4$  are incompatible with $E_{7(7)}$ symmetry
 because their $6$-point matrix elements have non-vanishing single-soft scalar limits.  Previous string theory~\ecite{\Eisenstein} and superspace~\cite{Bossard:2010bd} arguments suggested this $E_{7(7)}$-violation.
 Our results mean that no UV divergences  occur in $\cn =8$ supergravity below the 7-loop level.

 We then survey the candidate counter\-terms for loop orders
 $L = 7,8,9$ using two new algorithmic methods: one program counts monomials in the fields of $\cn=8$ supergravity in representations of the superalgebra $SU(2,2|8)$, the other applies Gr\"obner basis methods to  construct their explicit local matrix elements.
Our analysis shows that at each loop level 7, 8, 9, there is an infinite tower of independent $n$-point supersymmetric counterterms with $n \ge 4$. At $7$-loop order we find that none of the $n$-field operators with $n>4$ are  $E_{7(7)}$-compatible. This leaves $D^8R^4$ as the only candidate counterterm at $L=7$. We show that its matrix elements are $E_{7(7)}$-compatible at least up to $6$-points. We observe that it requires remarkable cancellations for $E_{7(7)}$  to be satisfied to all orders for any $L \ge 7$ candidate counterterm.

\section{\texorpdfstring{$E_{7(7)}$}{E7(7)}-violation of \texorpdfstring{$D^4R^4$}{D4R4} and \texorpdfstring{$D^6R^4$}{D6R4}}

To investigate $E_{7(7)}$ we study the soft scalar limit of the 6-point NMHV matrix elements  $\<\hp\hp\hm\hm\varphi\bar\varphi\>_{D^{2k}\!R^4}$. The external states are two pairs of opposite helicity gravitons and two conjugate scalars.
These matrix elements contain local terms from $n$th order field monomials in the nonlinear SUSY completion of $D^{2k}R^4$ as well as
non-local
pole diagrams in which one or more lines of the operator are off-shell and communicate to tree vertices from the classical Lagrangian. It is practically impossible to calculate these matrix elements with either Feynman rules (because the non-linear supersymmetrizations of $D^{2k}\!R^4$ are unknown) or recursion relations (because the matrix elements do not fall off under standard complex deformations of their external momenta).
Instead we use the $\alpha'$-expansion of the closed string tree amplitude to obtain the desired matrix elements.

At tree level, the closed string effective action takes the  form
\begin{eqnarray}
\label{effS}
     S_{\rm eff}&=&S_\text{SG}-2\alpha'^3\zeta(3)e^{\dash 6\phi}R^4-\zeta(5)\alpha'^5e^{\dash 10\phi}D^4R^4\\
     &&\!\!\!+\tfrac{2}{3}\alpha'^6\zeta(3)^2e^{\dash12\phi}D^6R^4
     -\tfrac{1}{2}\alpha'^7\zeta(7)e^{\dash14\phi}D^8R^4+\dots\,.
\nonumber
\end{eqnarray}
All closed string amplitudes in this work are obtained via KLT~\cite{KLT} from the open string amplitudes of \ecite{\StSt}. The amplitudes confirm the structure and coefficients of (\ref{effS}).

Couplings of the dilaton $\phi$ break the $SU(8)$-symmetry of the supergravity theory to $SU(4)\times SU(4)$ when $\alpha'>0$,
and thus the supersymmetric operators of $S_{\rm eff}$ are not the desired $SU(8)$-invariant operators. As explained in \cite{EK}, an $SU(8)$-averaging procedure extracts the $SU(8)$ singlet contribution from the string matrix elements. Specifically, the $SU(8)$ average of the $\<\hp\hp\hm\hm\varphi\,\bar\varphi\>_{e^{\dash(2k+6)\phi}D^{2k}\!R^4}$ matrix elements from string theory is
\begin{eqnarray}
 \label{su8avg}
 &&\<\hp\hp\hm\hm\varphi\bar\varphi\>_{\rm avg} = \tfrac{1}{35} \<\hp\hp\hm\hm\varphi^{1234}\varphi^{5678}\> \\[.5ex] \nonumber
 &&
 - \tfrac{16}{35} \<\hp\hp\hm\hm\varphi^{123|5}\varphi^{4|678}\>
 + \tfrac{18}{35} \<\hp\hp\hm\hm\varphi^{12|56}\varphi^{34|78}\>\,.
\end{eqnarray}
The 3 terms on the right side correspond to the 3 inequivalent ways to construct scalars from particles of the $\cn =4$ gauge theory, namely from gluons, gluinos, and $\cn=4$ scalars. There are 35 distinct embeddings of $SU(4)\times SU(4)$ in $SU(8)$. Averaging is sufficient to give the matrix elements of the $\cn =8$ field theory operator $R^4$, as done in \cite{EK}, and we extend it here to $D^4R^4$. For $D^6R^4$ a further correction is necessary and is discussed below.

Before proceeding, we note that the operators in the action~(\ref{effS}) are normalized such that their $4$-point matrix elements are  $\<\hp\hp\hm\hm\>=g(s,t,u)[12]^4\<34\>^4$ with
\begin{equation}\label{gs}
\begin{split}
    g_{R^4}&=1\,,\qquad\qquad\quad~\, g_{D^4R^4}=~s^2+t^2+u^2\,,\\
    g_{D^6R^4}&=s^3+t^3+u^3\,,~~\,     g_{D^8R^4}=(s^2+t^2+u^2)^2\,.
\end{split}
\end{equation}


\smallsection{5-loop counterterm \texorpdfstring{$D^4R^4$}{D4R4}}
At order $\alpha'^5$, the $SU(8)$-average (\ref{su8avg}) of the string theory
amplitudes directly  gives the matrix elements of the unique $SU(8)$-invariant supersymmetrization of $D^4R^4$.  The result is a complicated
 non-local expression, but its single-soft scalar limit is very simple and local, viz.
\begin{equation}\label{soft}
    \lim_{p_6\to 0}\<\hp\hp\hm\hm\varphi\bar\varphi\>_{D^4R^4}
    =-\tfrac{6}{7}[12]^4\<34\>^4\sum_{i<j}s_{ij}^2\,.
\end{equation}
Since this limit is non-vanishing, the operator $D^4R^4$ is incompatible with continuous $E_{7(7)}$ symmetry.

\smallsection{6-loop counterterm \texorpdfstring{$D^6R^4$}{D6R4}}
The single-soft scalar limit of the $SU(8)$-singlet part of the closed string matrix element at order $\alpha'^6$, obtained by $SU(8)$-averaging, is
\begin{equation}\label{softstringyD6R4}
    \lim_{p_6\to 0}\<\hp\hp\hm\hm\varphi\bar\varphi\>_{(e^{\dash12\phi}D^6\!R^4)_{\rm avg}}=-\tfrac{33}{35}[12]^4\<34\>^4\sum_{i<j}s_{ij}^3\,.
\end{equation}

It is important to realize that at order $\alpha'^6$, the
 $6$-point NMHV  closed string amplitudes receive contributions from diagrams involving one vertex from $e^{-12\phi}D^6R^4$ (together with vertices from the supergravity Lagrangian) \emph{and} from pole diagrams with two 4-point vertices of $e^{-6\phi}R^4$
 (which coincides with $R^4$ at $4$-points).
 Since
 $R^4$
  is not present in $\cn=8$ supergravity, its
contributions must be removed to extract the matrix elements of the supergravity operator $D^6R^4$.  The removal process must be supersymmetric.

We first compute the $R^4\!-\!R^4$ pole contributions to the 6-graviton NMHV matrix element   $\<\hm\hm\hm\hp\hp\hp\>$ as follows.  This amplitude has dimension 14.  Factorization at the pole determines the simple form
\begin{equation}\label{dandidit}
 \<12\>^4[45]^4\<3|P_{126}|6]^4/P_{126}^2 + 8 ~{\rm permutations}\,,
\end{equation}
up to a local polynomial.
The 9 terms correspond to the 9 distinct  3-particle pole diagrams.  The result~(\ref{dandidit}) is then checked by computation of the Feynman diagrams from the $R^4$ vertex \ecite{\grosssloan}.
 As the non-linear supersymmetrization of $R^4$ may contribute additional local terms, we also consider adding
the most general  gauge-invariant and bose-symmetric  polynomial of dimension $14$ that can contribute to $\<\hm\hm\hm\hp\hp\hp\>$, namely
\begin{equation}  \label{hediditagain}
\big(\<12\>\<23\>\<31\>[45][56][64]\big)^2 P_{123}^2\,.
\end{equation}

To incorporate SUSY, we separately show that there is a basis for  $SU(8)$-invariant 6-particle NMHV superamplitudes (an alternative to the basis in \cite{EFKWI}) consisting of  $\<\hm\hm\hm\hp\hp\hp\>$ and 8 distinct permutations of the states.  In this basis we write a superamplitude ansatz as the sum of the pole amplitude (\ref{dandidit}) plus a multiple of (\ref{hediditagain}).   We then impose full $S_6$ permutation symmetry on the ansatz.  This fixes the coefficient of the polynomial~(\ref{hediditagain}) to vanish and   determines the SUSY completion of the desired pole diagram uniquely!

Finally we project out the scalar-graviton matrix element from this superamplitude and take its single-soft scalar limit to find
\begin{equation}\label{mkdidit}
    \lim_{p_6\to 0}\<\hp\hp\hm\hm\varphi\bar\varphi\>_{(R^4)^2}=-\tfrac{1}{70}[12]^4\<34\>^4\sum_{i<j}s_{ij}^3\,.
\end{equation}
It is this contribution that we need to subtract from~(\ref{softstringyD6R4}) to obtain the single-soft scalar limits of the unique independent $D^6R^4$ operator in $\cn=8$ supergravity. Taking the relative normalization
$[-2\alpha'^3\zeta(3)]^2/[\tfrac{2}{3}\alpha'^6\zeta(3)^2] = 6$
of operators in the string effective action~(\ref{effS}) into account, we obtain
\begin{equation}\label{softD6R4}
    \lim_{p_6\to 0}\<\hp\hp\hm\hm\varphi\bar\varphi\>_{D^6\!R^4}=-\tfrac{30}{35}[12]^4\<34\>^4\sum_{i<j}s_{ij}^3\,.
\end{equation}
This non-vanishing result shows that the operator $D^6R^4$ is also incompatible
with continuous $E_{7(7)}$ symmetry.

$R^4$, $D^4R^4$ and $D^6R^4$ are the only local supersymmetric and $SU(8)$-symmetric operators for loop levels $L\leq 6$ \cite{Harmonic,EFK,NonHarmonic}. Hence $\cn=8$ supergravity has no potential counterterms that satisfy the continuous $E_{7(7)}$ symmetry for $L\leq 6$.  We stress that  string theory is used as a tool to extract $SU(8)$-invariant
 matrix elements that must agree with the
 matrix elements of the $\cn=8$ supergravity operators $R^4$, $D^4R^4$ and $D^6R^4$ because each of these operators is unique.
 No remnant of string-specific dynamics remains in the final results.

\smallsection{Matching to automorphism analysis}
The non-vanishing single-soft scalar limits found above have their origin in local 6-point interactions of the schematic form $\varphi^2 D^{2k}\!R^4$ which appear in the non-linear completion of $D^{2k}R^4$. Let us encode this completion as $f(\varphi)D^{2k}R^4$, with
\begin{equation}
    f(\varphi)=1-a \bigl[\varphi^{1234}\varphi^{5678}+34\text{ inequiv. perms}\bigr]+\ldots ~\,.
    \label{f}
\end{equation}
The ``\dots'' indicate higher order terms.
The constant $a$ depends on the operator; for example  $a_{R^4}=\tfrac{6}{5}$ for $R^4$ \cite{EK}. We can determine $a$ for $D^{4}R^4$ by taking a further single-soft limit $p_5\!\to\!0$ on~(\ref{soft}) and comparing the resulting $s,t,u$-polynomial with the $4$-point normalization of~(\ref{gs}). The result is  $a_{D^{4}\!R^4}=\tfrac{12}{7}$.

Ref.~\ecite{\Eisenstein} used supersymmetry and duality considerations in $d$ dimensions to constrain the moduli dependent functions $f(\varphi)$ of the BPS operators $R^4$, $D^{4}R^4$ and $D^{6}R^4$. Specifically, for $R^4$ and $D^4R^4$ in $4$ dimensions, they found that $f(\varphi)$ should satisfy the Laplace equation
\begin{equation}\label{LapD4R4}
    (\Delta+42) f_{R^4}(\varphi)=0\,,\quad (\Delta+60) f_{D^4R^4}(\varphi)=0\,.
\end{equation}
Here, $\Delta$ is the Laplacian on $E_{7(7)}/SU(8)$; in terms of the scalars $\varphi^{abcd}$ of $\cn=8$ supergravity, its leading terms are
\begin{equation}
  \Delta =
  \Big[\frac{\partial}{\partial\varphi^{1234}} \frac{\partial}{\partial\varphi^{5678}}
  +
  ~\text{34 inequiv. perms}\Big] + \dots~\,.
\end{equation}
It is easy to see that  the function
(\ref{f}) with the above values of $a$ precisely satisfy the Laplace equations~(\ref{LapD4R4}). This is a consistency check on our result for the   single-soft scalar limits.

Let us now consider the function $f(\varphi)$ for $D^6R^4$. As explained, the quadratic order of $R^4$ interferes with $D^6R^4$ and it is therefore natural that the corresponding Laplace equation in~\ecite{\Eisenstein} contains an inhomogeneous term that reflects the contribution from $R^4\!-\!R^4$. Adding a general linear combination $\lambda_{R^4} f_{R^4}R^4+\lambda_{D^6\!R^4} f_{D^6R^4}D^6\!R^4$ to the effective action constrains the moduli-dependent functions to satisfy~\ecite{\Eisenstein}
\begin{equation}
\begin{split}\label{LapD6R4}
    (\Delta+60) f_{D^6R^4}(\varphi)&=-\frac{\lambda_{R^4}^2}{\lambda_{D^6\!R^4}}\bigl[f_{R^4}(\varphi)\bigr]^2\,.
\end{split}
\end{equation}
From~(\ref{softstringyD6R4}) and~(\ref{softD6R4}), we can reconstruct the coefficient $a$ in~(\ref{f}) of the functions associated with the $SU(8)$-averaged string-theory operator $(e^{\dash12\phi}D^6\!R^4)_{\rm avg}$ and with the supergravity operator $D^6R^4$. We find
\begin{equation}
        a_{(e^{\dash12\phi}D^6\!R^4)_{\rm avg}}=\tfrac{66}{35}\,,\qquad
        a_{D^6\!R^4}=\tfrac{60}{35}\,.
\end{equation}
For $(e^{\dash12\phi}D^6\!R^4)_{\rm avg}$, the couplings $\lambda$ in~(\ref{LapD6R4}) must take their string theory values $\lambda_{R^4}\!=\!-2\alpha'^3\zeta(3)$ and
$\lambda_{D^6\!R^4}\!=\!\frac{2}{3}\alpha'^6\zeta(3)^2$. The $\cn=8$  operator $D^6R^4$, on the other hand, must satisfy~(\ref{LapD6R4}) with $\lambda_{R^4}=0$ because the operator $R^4$ does not appear in the action of  $\cn=8$ supergravity.  Indeed, our results for $f$ for both operators satisfy the Laplace
equation with the expected choice of $\lambda$'s.

\section{Construction and counting of counterterms}
We now discuss the techniques used to classify
and  construct  local supersymmetric operators, especially those needed for $L\ge 7$. We are interested in $SU(8)$-invariant operators, which are candidate counter\-terms, and in operators transforming in the ${\bf 70}$ of $SU(8)$. The latter are candidate operators for local single-soft scalar limits (SSL's) of the matrix elements of singlet counter\-term operators. First we use representation theory of the  superalgebra $SU(2,2|8)$ to determine the spectrum and multiplicity of these operators.  The spectrum is classified by the number $n$ of external fields,  the scale dimension, and the order $k$ of the N$^k$MHV type. Then we construct  matrix elements of several operators explicitly using algorithms  which incorporate Gr\"obner basis techniques.

\smallsection{Spectrum of local operators}
Counterterms of $\cn=8$ supergravity are
supersymmetric,
$SU(8)$-invariant,
Lorentz scalar
local operators $C$ integrated over  spacetime.
These local operators involve $n$-fold products
of the fundamental fields and their derivatives. We restrict to diffeomorphism-covariant  combinations of the fields, such as the Riemann tensor $R$.
In enumerating all local operators of a given order $n$ (up to covariance), the equations of motion set the Ricci tensor equal to a combination of fields of quadratic order (and higher), which is automatically included at order  $> n$ in the enumeration.
The remaining 10 components of the on-shell Riemann tensor group into
fields with Lorentz spin $(2,0)$ and $(0,2)$.  The collection of
 all on-shell supergravity fields span  a
representation of the $\cn=8$ super-Poincar\'e algebra as well as
an ultrashort representation of $\cn=8$ superconformal symmetry
(see \cite{NonHarmonic} for a recent discussion).
Using the $SU(2,2|8)$ Dynkin diagram
\begin{equation}
\mathnormal{\mathop{\mathnormal{\odot}}_{\makebox[0pt][r]{\footnotesize$SU(2)_{\mathrm{L}}$}}}\text{--}
\mathnormal{\otimes}\text{--}
\mathnormal{\underbrace{\mathnormal{\odot}\text{--}
\mathnormal{\odot}\text{--}
\mathnormal{\odot}\text{--}
\mathnormal{\odot}\text{--}
\mathnormal{\odot}\text{--}
\mathnormal{\odot}\text{--}
\mathnormal{\odot}}_{\text{\footnotesize$SU(8)$}}}\text{--}
\mathnormal{\otimes}\text{--}
\mathnormal{\mathop{\mathnormal{\odot}}_{\makebox[0pt][l]{~~\footnotesize$SU(2)_{\mathrm{R}}$}}}\,\qquad,
\end{equation}
the Dynkin labels of this lowest-weight representation read $[0\mathord{,}0\mathord{,}0001000\mathord{,}0\mathord{,}0]$, where the $SU(8)$ labels $[0001000]$ describe a $\textbf{70}$
and the $SU(2)_{\mathrm{L}}\times SU(2)_{\mathrm{R}}$ Lorentz spins indicate a scalar.

The graded symmetric tensor product of $n$ copies of the above multiplet
provide all local operators with $n$ fields.
We are interested in supersymmetric operators:
there is typically one such operator $C$ in each irrep of the tensor product.
For long supermultiplets it is the unique top component,
obtained by acting with SUSY generators  $Q^{16}\tilde Q^{16}$ on
the lowest-weight component $C_0$.
(In superspace approaches, this is equivalent to the full superspace measure $\int d^{32}\theta$.)
For short or BPS supermultiplets fewer supersymmetries are needed to get
from $C_0$ to the top component(s). Hence it is sufficient to enumerate the lowest superconformal weights $C_0$.
Its superconformal transformation properties determine the spin, $SU(8)$ representation
as well as loop and N$^k$MHV level.

More concretely, Dynkin labels translate to scalar local operators
as follows
(assume $q\geq p$)
\begin{equation}
\begin{array}{rcl}
[0\mathord{,}p\mathord{,}0000000\mathord{,}q\mathord{,}0]
\mathrel{}&\hspace{-2\arraycolsep}\to\hspace{-2\arraycolsep}&\mathrel{}
D^{3p-q-n}\varphi^{n+p-q} R^{q-p}\ \,\,\,\,(\mathbf{singlet})\,,
\\[0.5ex]
[0\mathord{,}p\mathord{,}0001000\mathord{,}q\mathord{,}0]
\mathrel{}&\hspace{-2\arraycolsep}\to\hspace{-2\arraycolsep}&\mathrel{}
D^{3p-q-n+1}\varphi^{n+p-q} R^{q-p}\ \,\,\,\,(\mathbf{70})\,.
\end{array}
\label{npq}
\end{equation}
Note that we display only prototypical terms,
mixture with other fields is implied:
E.g.~$D^4\varphi^2\simeq R\bar{R}$. (Here we distinguish between the chiral and anti-chiral components of the Riemann tensor.)
To get from $C_0$ to the supersymmetric $C$ in long multiplets,
apply $Q^{16}\tilde Q^{16}\simeq D^{16}$.
A lowest weight as in \eqref{npq} then
corresponds to a N$^k$MHV counterterm at $L$ loops with
\begin{equation}
\label{kL}
2k=n+p-q-4\,,\hspace{3mm}
2L=\Big\{
\begin{array}{lc}
14+p+q-n & ({\bf singlet}),\\
15+p+q-n & ({\bf 70}).
\end{array}
\end{equation}

Note that locality requires that the exponents in (\ref{npq}) are non-negative numbers. In particular, $3p-q-n \ge 0$ implies $2L\ge 6+2n - 4k$ using (\ref{kL}). This bound on the existence of local non-BPS operators was conjectured in \cite{EFK} and confirmed very recently in \cite{NonHarmonic}.

As a simple illustration, consider (\ref{npq}) with $n=4$ and $p=q$. We find $C_0 = D^{2p-4} \varphi^4$, and after application of $D^{16}$ it becomes $D^{2p+4} R^2 \bar{R}^2$; this is just the 4-point MHV local counter\-term $D^{2p+4}R^4$. Locality of $C_0$ requires $p \ge 2$, so the first available non-BPS operator is $D^8R^4$.

In practice, we use a \texttt{C++} program to enumerate all local operators
with $2L\le 30-n$ amounting to $\sim 4.8\cdot 10^{22}$ terms.\footnote{The computation took 3.5 hours on a desktop PC.}
These are decomposed into irreps of $SU(2,2|8)$
by iteratively removing the lowest weights
and their corresponding supermultiplets \ecite{\NF}.
Special attention needs to be paid to BPS and short supermultiplets \ecite{\shorts}.
In total we obtained around $8.8\cdot 10^5$ types of supermultiplets along with
their multiplicities.\footnote{The decomposition took 42 hours.}
Finally we extract supermultiplets with
scalar $SU(8)$ singlets and $\textbf{70}$'s as top supersymmetry components.
The results at $L\leq 9$ are presented in Table~\ref{ourfigure}.

Our analysis shows that there are unique $\frac{1}{2}$, $\frac{1}{4}$, $\frac{1}{8}$
BPS counterterms $R^4$, $D^4R^4$ and $D^6R^4$, in agreement with earlier results \cite{Harmonic,EFK,NonHarmonic}. They correspond to the lowest weights
\begin{equation}
[0\mathord{,}0\mathord{,}0004000\mathord{,}0\mathord{,}0],\
[0\mathord{,}0\mathord{,}0200020\mathord{,}0\mathord{,}0],\
[0\mathord{,}0\mathord{,}2000002\mathord{,}0\mathord{,}0]\,.
\end{equation}
In the previous section, we showed that their 6-point matrix elements have non-vanishing single-soft limits originating from the non-linear completion of the operators. The limits correspond to local ${\bf 70}$ BPS operators $\varphi R^4$, $\varphi D^4R^4$ and $\varphi D^6R^4$, which are descendants of the
$\frac{1}{2}$, $\frac{1}{4}$, $\frac{1}{8}$ BPS
superconformal primaries $\varphi^5$
with $SU(8)$ Dynkin labels $[0005000]$, $[0201020]$, $[2001002]$. The relationship between BPS operators are illustrated in Table~\ref{ourfigure}.


\smallsection{Explicit matrix elements and superamplitudes}
The matrix elements of potential counterterms such as $D^{2k}R^n$ must be polynomials of degree $\delta = 2(k+n)$ in angle and square brackets $\<i\,j\>,~ [kl]$ which satisfy several constraints.  If $a_i$ and $s_i$ denote the number of angle $|i\>$ and square $|i]$ spinors for
each particle $i=1,2,\dots,n$, then the total number of spinors is fixed by the dimension of the operator to be $\sum_i(a_i+s_i) = 4(k+n)$.  For each particle $i$, of helicity $h_i$,  there is a helicity weight constraint $a_i - s_i = - 2h_i$.  We need polynomials which are independent under the constraints of momentum  conservation and the Schouten identity,
\begin{equation} \label{ideal}
    \sum_j\<ij\>[jk]=0\,,\quad
    \<ij\>\<kl\>+\<jk\>\<il\>+\<ki\>\<jl\>=0\,,
\end{equation}
and a similar Schouten identity for square brackets. These polynomials must satisfy  bose and fermi symmetries when they contain identical particles.  Finally, the polynomials must satisfy SUSY Ward identities. This
 was ensured in \cite{EFK} by packaging $n$-point matrix elements into the manifestly SUSY- and $SU(8)$-invariant superamplitudes of \cite{EFKWI}.

In \cite{EFK}, Mathematica was used to construct the required independent polynomials.
More efficient algorithms are needed for  the higher dimension counter\-terms studied in this paper.
 The constraints (\ref{ideal}) define an ideal in a polynomial ring, and the Gr\"obner basis method  \cite{clo,bstru}
 is well  suited to choose a basis in the ideal and generate independent sets of polynomials in the quotient ring.

Given a (conventional) monomial ordering in the ring, a Gr\"obner basis is
a subset of the ideal such that the  leading term of any element of the
ideal is divisible by a leading term of an element of the subset.
Buchberger's algorithm generates the unique reduced Gr\"obner basis in
which no monomial in a polynomial of this  basis is divisible by a leading
term of the other polynomials in the basis. For the ideal generated by
(\ref{ideal}), the reduced Gr\"obner basis is quite simple.  By the theory
of Gr\"obner bases, the monomials of degree $\delta$ (and specific
helicity weights) that are not divisible by any leading term of the reduced
Gr\"obner basis are a vector space basis of the quotient ring.    This
division test concerns only monomials and is computationally  fast.
(See Ch.~2, Sec.~7  and Prop.~5.3.1 of  \cite{clo}.)

We used the implementation of the Buchberger's algorithm in the algebraic
software system Macaulay2 \cite{Mac2} to generate independent
polynomials which satisfy dimension and helicity weight requirements and
the constraints (\ref{ideal}).  These polynomials were then processed by
computer programs similar to those used  in \cite{EFK} which imposed bose
symmetries. Among the resulting polynomials we select the ones that are independent under the conditions~(\ref{ideal}).

We have applied the Gr\"obner basis method to local counterterms with $n\le6$.
The results are in perfect agreement with the multiplicities found from the enumeration of $SU(2,2|8)$ superconformal primary operators. In addition to an enumeration of independent operators, the explicit matrix elements allow us to test single-soft scalar limits. We discuss our $L=7,8,9$ results below.


\begin{table*}[t!]
\footnotesize{
%
%
\begin{tabular}{|c|ccc|}
\hline
{\bf 3-loop}&4-pt&\!\!\!5-pt\,&6-pt\\
\hline
&&&\\[-10pt]
{\bf singlet}\,&
\begin{tabular}{c}
$\bdgreen R^4$  \\[-0.5mm]
{\tiny \bdgreen  $1\times$MHV}
\end{tabular}
&
\!\!\!
\begin{tabular}{c}
$\bdr \cancel{\varphi^2  D^2\!R^3}$\\
\,{\scriptsize \phantom{$D^4R^4$}}
\end{tabular}
\,
&
\hspace{-4mm}
\begin{tabular}{c}
$~~~\bdr \cancel{\varphi^2  R^4}$  \\[0.2mm]
\,{\scriptsize \bdgreen  $R^4$ non-linear}\!\!
\end{tabular}
\\[-0.1mm]
&&&\hspace{-18mm}$\bdb \swarrow_{\rm soft}$\\[1mm]
{\bf 70}\,
&
&
\hspace{-4.5mm}
\begin{tabular}{c}
$\bdb 1\!\times\! \varphi R^4$
\end{tabular}
&~\\[.9mm]
\hline
\end{tabular}
\hspace{2mm}
%
%
\begin{tabular}{|c|ccc|}
\hline
{\bf 5-loop}&4-pt&\!\!\!5-pt\,&6-pt\\
\hline
&&&\\[-10pt]
{\bf singlet}\,&
\begin{tabular}{c}
$\bdgreen D^4R^4$  \\[-0.5mm]
{\tiny \bdgreen  $1\times$MHV}
\end{tabular}
&
\!\!
\begin{tabular}{c}
$\bdr \cancel{\varphi^2  D^6\!R^3}$\\
{\scriptsize \phantom{$D^4\!R^4$}}
\end{tabular}
\,\,
&
\hspace{-4mm}
\begin{tabular}{c}
~~~$\bdr \cancel{\varphi^2  D^4R^4}$  \\[0.2mm]
\,{\scriptsize \bdgreen  $D^4\!R^4$ non-lin.}\!\!
\end{tabular}
\\[-0.1mm]
&&&\hspace{-18mm}$\bdb \swarrow_{\rm soft}$\\[1mm]
{\bf 70}\,
&
&
\hspace{-4.5mm}
\begin{tabular}{c}
$\,\bdb 1\!\times\! \varphi D^4\!R^4$
\end{tabular}
&~\\[.9mm]
\hline
\end{tabular}
\hspace{2mm}
%
%
\begin{tabular}{|c|ccc|}
\hline
{\bf 6-loop}&4-pt&\!5-pt\,&6-pt\\
\hline
&&&\\[-10pt]
{\bf singlet}\,&
\begin{tabular}{c}
$\bdgreen D^6R^4$  \\[-0.5mm]
{\tiny \bdgreen  $1\times$MHV}
\end{tabular}
&
\!\!\!
\begin{tabular}{c}
~~~$\bdr \cancel{\varphi^2  D^8\!R^3}$\\
\,{\scriptsize \phantom{$D^4R^4$}}
\end{tabular}
\,\,
&
\hspace{-4mm}
\begin{tabular}{c}
$\bdr \cancel{\varphi^2  D^6R^4}$  \\[0.2mm]
\,{\scriptsize \bdgreen  $D^6R^4$ non-lin.}\!
\end{tabular}
\\[-0.1mm]
&&&\hspace{-18mm}$\bdb \swarrow_{\rm soft}$\\[1mm]
{\bf 70}\,
&
&
\hspace{-4.5mm}
\begin{tabular}{c}
$\bdb 1\!\times\! \varphi D^6R^4$
\end{tabular}
&~\\[.9mm]
\hline
\end{tabular}
\\[2mm]
%
%
\begin{tabular}{|c|ccccccccccccc|}
\hline
{\bf 7-loop}&4-pt&5-pt&6-pt&7-pt&8-pt&9-pt&10-pt&11-pt&12-pt&13-pt&14-pt&15-pt&16-pt\\
\hline
&&&&&&&&&&&&&\\[-8pt]
{\bf singlet}\,
&
\hspace{-1mm}
\begin{tabular}{c}
$\bdgreen D^8\!  R^4$  \\[-1mm]
{\tiny \bdgreen  $1\times$MHV}
\end{tabular}
&
\hspace{-2mm}
\begin{tabular}{c}
$\bdr \cancel{D^6\!  R^5}$  \\[-1mm]
\phantom{\tiny \bdr  $\nexists$}
\end{tabular}
\hspace{-2mm}
&
\hspace{-2mm}
\begin{tabular}{c}
$\bdgreen D^4\!  R^6$  \\[-1mm]
{\tiny \bdgreen 2$\times$NMHV}
\end{tabular}
\hspace{-2mm}
&
\hspace{-2mm}
\begin{tabular}{c}
$\bdr \cancel{D^2\!  R^7}$  \\[-1mm]
\phantom{\tiny \bdr  $\nexists$}
\end{tabular}
\hspace{-2mm}
&
\hspace{-2mm}
\begin{tabular}{c}
$\bdgreen R^8$  \\[-1mm]
{\tiny \bdgreen $3\times$N$^2$MHV}
\end{tabular}
\hspace{-2mm}
&
\hspace{-1mm}
\begin{tabular}{c}
$\bdr \cancel{\varphi^2 D^2\!  R^7}$  \\[-1mm]
\phantom{\tiny \bdr  $\nexists$}
\end{tabular}
\hspace{-1mm}
&
\hspace{-2mm}
\begin{tabular}{c}
$\bdgreen \varphi^2 R^8$ \\[-1mm]
{\tiny \bdgreen $4\times$N$^3$MHV}
\end{tabular}
\hspace{-2mm}
&
\hspace{-1mm}
\begin{tabular}{c}
$\bdr \cancel{\varphi^4 D^2\!  R^7}$  \\[-1mm]
\phantom{\tiny \bdr  $\nexists$}
\end{tabular}
\hspace{-1mm}
&
\hspace{-2mm}
\begin{tabular}{c}
$\bdgreen \varphi^4 R^8$  \\[-1mm]
{\tiny \bdgreen $6\times$N$^4$MHV}
\end{tabular}
\hspace{-2mm}
&
\hspace{-1mm}
\begin{tabular}{c}
$\bdr \cancel{\varphi^6 D^2\!  R^7}$  \\[-1mm]
\phantom{\tiny \bdr  $\nexists$}
\end{tabular}
\hspace{-1mm}
&
\hspace{-2mm}
\begin{tabular}{c}
$\bdgreen \varphi^6 R^8$  \\[-1mm]
{\tiny \bdgreen $8\times$N$^5$MHV}
\end{tabular}
\hspace{-2mm}
&
\hspace{-1mm}
\begin{tabular}{c}
$\bdr \cancel{\varphi^8 D^2\!  R^7}$  \\[-1mm]
\phantom{\tiny \bdr  $\nexists$}
\end{tabular}
\hspace{-1mm}
&
\hspace{-2mm}
\begin{tabular}{c}
$\bdgreen \varphi^8 R^8$  \\[-1mm]
{\tiny \bdgreen $10\times$N$^6$MHV}
\end{tabular}
\hspace{-2mm}
\\[1mm]
&
&\hspace{4mm}$\blb \swarrow_{\rm soft}$\hspace{-9mm}
&&\hspace{5mm}$\blb \swarrow_{\rm soft}$\hspace{-9mm}
&&\hspace{4mm}$\blb \swarrow_{\rm soft}$\hspace{-9mm}
&&\hspace{6.5mm}$\blb \swarrow_{\rm soft}$\hspace{-9mm}
&&\hspace{8mm}$\blb \swarrow_{\rm soft}$\hspace{-9mm}
&&\hspace{8mm}$\blb \swarrow_{\rm soft}$\hspace{-9mm}
&~
\\[0.5mm]
\,${\bf 70}$\,
&
&
\hspace{-2mm}
\begin{tabular}{c}
$\blb \varphi D^8 \!R^4$  \\[-1.3mm]
{\tiny \blb  $2\!\times$
}
\end{tabular}
\hspace{0mm}
&
&
\hspace{-1mm}
\begin{tabular}{c}
$\blb \varphi D^4 \!R^6$  \\[-1.3mm]
{\tiny \blb  $4\times$}
\end{tabular}
\hspace{-1mm}
&
&
\begin{tabular}{c}
$\blb \varphi R^8$  \\[-1.3mm]
{\tiny \blb  $6\times$}
\end{tabular}
&
&
\begin{tabular}{c}
$\blb \varphi^3R^8$  \\[-1.3mm]
{\tiny \blb $9\times$}
\end{tabular}
&
&
\begin{tabular}{c}
$\blb  \varphi^5 R^8$  \\[-1.3mm]
{\tiny \blb  $14\times$}
\end{tabular}
&
&
\begin{tabular}{c}
$\blb  \varphi^7 R^8$  \\[-1.3mm]
{\tiny \blb $19\times$}
\end{tabular}
&~\\[1mm]
\hline
\end{tabular}%
\vspace{2mm}\par
%
%
\begin{tabular}{|c|cccccccccccc|}
\hline
{\bf 8-loop}&4-pt&5-pt&6-pt&7-pt&8-pt&9-pt&10-pt&11-pt&12-pt&13-pt&14-pt&\\
\hline
&&&&&&&&&&&&\\[-8pt]
{\bf singlet}\,
&
\begin{tabular}{c}
$\bdgreen D^{10}\!  R^4$ \\[-1mm]
{\tiny \bdgreen $1 \times$MHV}
\end{tabular}
&
\hspace{-0.8mm}
\begin{tabular}{c}
$\bdgreen D^8\!  R^5$\\[-1mm]
{\tiny \bdgreen  1$\times$MHV}
\end{tabular}
&
\hspace{-1mm}
\begin{tabular}{c}
$\bdgreen D^6\!  R^6$ \\[-1mm]
{\tiny \bdgreen $3\times$NMHV}
\end{tabular}
&
\hspace{.5mm}
\begin{tabular}{c}
$\bdgreen D^4\!  R^7$\\[-1mm]
{\tiny \bdgreen $3\times$NMHV}
\end{tabular}
\hspace{.5mm}
&
\hspace{-1mm}
\begin{tabular}{c}
$\bdgreen D^2\!  R^8$ \\[-1mm]
{\tiny \bdgreen $8\times$N$^2$MHV}
\end{tabular}
&
\hspace{-1mm}
\begin{tabular}{c}
$\bdgreen R^9$ \\[-1mm]
{\tiny \bdgreen $8\times$N$^2$MHV}
\end{tabular}
&
\hspace{-1mm}
\begin{tabular}{c}
$\bdgreen \varphi^2 D^2\!  R^8$ \\[-1mm]
{\tiny \bdgreen $25\times$N$^3$MHV}
\end{tabular}
&
\hspace{-1mm}
\begin{tabular}{c}
$\bdgreen \varphi^2 R^9$ \\[-1mm]
{\tiny \bdgreen $22\times$N$^3$MHV}
\end{tabular}
&
\hspace{-1mm}
\begin{tabular}{c}
$\bdgreen \varphi^4\! D^2\!  R^8$ \\[-1mm]
{\tiny \bdgreen $66\times$N$^4$MHV}
\end{tabular}
&
\hspace{-1mm}
\begin{tabular}{c}
$\bdgreen \varphi^4\!  R^9$ \\[-1mm]
{\tiny \bdgreen $51\times$N$^4$MHV}
\end{tabular}
&
\hspace{-.75mm}
\begin{tabular}{c}
$\bdgreen \varphi^6\! D^2\!  R^8$ \\[-1mm]
{\tiny \bdgreen $153\times$N$^5$MHV}
\end{tabular}
\hspace{-.5mm}
&
\!\!\!
\\[1mm]
&
&\hspace{7mm}$\blb \swarrow$\hspace{-6mm}
&\hspace{8mm}$\blb \swarrow$\hspace{-6mm}
&\hspace{8mm}$\blb \swarrow$\hspace{-6mm}
&\hspace{8mm}$\blb \swarrow$\hspace{-6mm}
&\hspace{8mm}$\blb \swarrow$\hspace{-6mm}
&\hspace{5.7mm}$\blb \swarrow$\hspace{-6mm}
&\hspace{8mm}$\blb \swarrow$\hspace{-6mm}
&\hspace{8mm}$\blb \swarrow$\hspace{-6mm}
&\hspace{12.5mm}$\blb \swarrow$\hspace{-6mm}
&&~\!\!\!\!\!\!\!
\\[0.5mm]
\,${\bf 70}$\,
&
&
\hspace{-2mm}
\begin{tabular}{c}
$\blb \varphi D^{10}\! R^4$\\[-1.3mm]
{\tiny \blb  $3\!\times$}
\end{tabular}
&
\hspace{-1.5mm}
\begin{tabular}{c}
$\blb \varphi D^8\!  R^5$ \\[-1.3mm]
{\tiny \blb  $4\!\times$}
\end{tabular}
&
\hspace{-1.5mm}
\begin{tabular}{c}
$\blb \varphi D^6\!  R^6$ \\[-1.3mm]
{\tiny \blb  $17\times$}
\end{tabular}
&
\hspace{-3mm}
\begin{tabular}{c}
$\blb \varphi D^4\!  R^7$ \\[-1.3mm]
{\tiny \blb  $16\times$}
\end{tabular}
&
\hspace{-3mm}
\begin{tabular}{c}
$\blb \varphi D^2\!  R^8$\\[-1.3mm]
{\tiny \blb  $81\times$}
\end{tabular}
&
\hspace{-0mm}
\begin{tabular}{c}
$\blb \varphi R^9$ \\[-1.3mm]
{\tiny \blb  $63\times$}
\end{tabular}
&
\hspace{-1.3mm}
\!\!\!\!
\begin{tabular}{c}
$\blb \varphi^3 D^2\!  R^8$\\[-1.3mm]
{\tiny \blb  $232\times$}
\end{tabular}
\!\!\!
&
\hspace{-1.5mm}
\begin{tabular}{c}
$\blb \varphi^3 R^9$ \\[-1.3mm]
{\tiny \blb  $211\times$}
\end{tabular}
&
\hspace{-0mm}
\!\!\!\!\!
\begin{tabular}{c}
$\blb \varphi^5 D^2\!  R^8$\\[-1.3mm]
{\tiny \blb  $1033\times$}
\end{tabular}
\!\!\!\!\!
&&\\[1mm]
\hline
\end{tabular}%
\vspace{2mm}\par
%
%
%
%
%
%
%
\begin{tabular}{|c|cccccccccc|}
\hline
{\bf 9-loop}&4-pt&5-pt&6-pt&7-pt&8-pt&9-pt&10-pt&11-pt&12-pt&\\
\hline
&&&&&&&&&&\\[-8pt]
{\bf singlet}\,
&
\begin{tabular}{c}
$\bdgreen D^{12}\!  R^4$ \\[-1mm]
{\tiny \bdgreen $2 \times$MHV}\\[-2mm]
{\tiny \phantom{MHV}}
\end{tabular}
&
\hspace{-0.8mm}
\begin{tabular}{c}
$\bdgreen D^{10}\!  R^5$\\[-1mm]
{\tiny \bdgreen  1$\times$MHV}\\[-2mm]
{\tiny \phantom{MHV}}
\end{tabular}
&
\hspace{-1mm}
\begin{tabular}{c}
$\bdgreen D^8\!  R^6$ \\[-1mm]
{\tiny \bdgreen $12\times$NMHV}\\[-2mm]
{\tiny \bdgreen $2\times$MHV\,}
\end{tabular}
&
\hspace{2mm}
\begin{tabular}{c}
$\bdgreen D^6\!  R^7$\\[-1mm]
{\tiny \bdgreen $14\times$NMHV}\\[-2mm]
{\tiny \phantom{MHV}}
\end{tabular}
\hspace{3mm}
&
\hspace{-1mm}
\begin{tabular}{c}
$\bdgreen D^4\!  R^8$ \\[-1mm]
{\tiny \bdgreen $117\times$N$^2$MHV}\\[-2mm]
{\tiny \bdgreen ~$7\times$NMHV}
\end{tabular}
\hspace{2mm}
&
\hspace{-3mm}
\begin{tabular}{c}
$\bdgreen D^2\! R^9$ \\[-1mm]
{\tiny \bdgreen $123\times$N$^2$MHV}\\[-2mm]
{\tiny \phantom{MHV}}
\end{tabular}
&
\hspace{-1mm}
\begin{tabular}{c}
$\bdgreen R^{10}$ \\[-1mm]
{\tiny \bdgreen $780\times$N$^3$MHV}\\[-2mm]
{\tiny \bdgreen ~\,$36\times$N$^2$MHV}
\end{tabular}
&
\hspace{-1mm}
\begin{tabular}{c}
$\bdgreen \varphi^2 D^2\!R^9$ \\[-1mm]
{\tiny \bdgreen $783\times$N$^3$MHV}\\[-2mm]
{\tiny \phantom{MHV}}
\end{tabular}
&
\hspace{-1mm}
\begin{tabular}{c}
$\bdgreen \varphi^2  R^{10}$ \\[-1mm]
{\tiny \bdgreen $4349\times$N$^4$MHV}\\[-2mm]
{\tiny \bdgreen ~\,$169\times$N$^3$MHV}
\end{tabular}
\hspace{-2.2mm}
&
\!\!\!
\\[1mm]
&
&\hspace{7mm}$\blb \swarrow$\hspace{-6mm}
&\hspace{8mm}$\blb \swarrow$\hspace{-6mm}
&\hspace{8mm}$\blb \swarrow$\hspace{-6mm}
&\hspace{10mm}$\blb \swarrow$\hspace{-6mm}
&\hspace{10mm}$\blb \swarrow$\hspace{-6mm}
&\hspace{15.7mm}$\blb \swarrow$\hspace{-6mm}
&\hspace{12mm}$\blb \swarrow$\hspace{-6mm}
&&~\!\!\!\!\!\!\!
\\[0.5mm]
\,${\bf 70}$\,
&
&
\hspace{-2mm}
\begin{tabular}{c}
$\blb \varphi D^{12}\! R^4$\\[-1.3mm]
{\tiny \blb  $5\times$N$^{0.5}$MHV}\\[-1.5mm]
{\tiny \phantom{MHV}}
\end{tabular}
\hspace{1mm}
&
\hspace{-1.5mm}
\begin{tabular}{c}
$\blb \varphi D^{10}\!  R^5$ \\[-1.3mm]
{\tiny \blb  $8\times$N$^{0.5}$MHV}\\[-1.5mm]
{\tiny \phantom{MHV}}
\end{tabular}
&
\hspace{-1.5mm}
\begin{tabular}{c}
$\blb \varphi D^8\!  R^6$ \\[-1.3mm]
{\tiny \blb  $122\times$N$^{1.5}$MHV}\\[-1.5mm]
{\tiny \blb ~~\,$5\times$N$^{0.5}$MHV}
\end{tabular}
&
\hspace{-3mm}
\begin{tabular}{c}
$\blb \varphi D^6\!  R^7$ \\[-1.3mm]
{\tiny \blb  $194\times$N$^{1.5}$MHV}\\[-1.5mm]
{\tiny \phantom{MHV}}
\end{tabular}
\hspace{-1mm}
&
\hspace{-3mm}
\begin{tabular}{c}
$\blb \varphi D^4\!  R^8$\\[-1.3mm]
{\tiny \blb  $1814\times$N$^{2.5}$MHV}\\[-1.5mm]
{\tiny \blb ~~\,$52\times$N$^{1.5}$MHV}
\end{tabular}
&
\hspace{-0mm}
\begin{tabular}{c}
$\blb \varphi D^2\!  R^9$ \\[-1.3mm]
{\tiny \blb  $2317\times$N$^{2.5}$MHV}\\[-1.5mm]
{\tiny \phantom{MHV}}
\end{tabular}
&
\hspace{.3mm}
\!\!\!\!
\begin{tabular}{c}
$\blb \varphi R^{10}$\\[-1.3mm]
{\tiny \blb  $16485\times$N$^{3.5}$MHV}\\[-1.5mm]
{\tiny \blb ~~\,$469\times$N$^{2.5}$MHV}
\end{tabular}
\!\!\!\!\!
&&\\[1mm]
\hline
\end{tabular}\par
}%
\caption{Supersymmetric $SU(8)$-singlet $L$-loop counterterms and the $SU(8)$ ${\bf 70}$ operators which describe their potential single-soft scalar limits. When the singlet operator is in the N$^k$MHV classification, the single-soft scalar limit operator belongs to the N$^{(k-\frac{1}{2})}$MHV sector. For $L<7$, there are no independent singlet operators with $n>4$, but the non-vanishing single-soft scalar limits arise from the non-linear completions of the 4-point operators $D^{2k}R^4$.
}
\label{ourfigure}
\end{table*}

\section{7-loop counterterms: \texorpdfstring{$D^8R^4$}{D8R4} and beyond}
\smallsection{\texorpdfstring{$E_{7(7)}$}{E7(7)}-compatibility of \texorpdfstring{$D^8R^4$}{D8R4} at 6-points}
The 6-point closed-string tree amplitude at order $\alpha'^7$ only receives contributions from diagrams with one insertion of
$e^{\dash14\phi}\!D^8\!R^4$. No lower-dimension operators in the closed string effective action (\ref{effS}) contribute. For the $SU(8)$-averaged single-soft scalar limits of $e^{\dash14\phi}D^8R^4$ we obtain
\begin{equation}\label{softstringyD8R4}
 \begin{split}
    &\lim_{p_6\to 0}\<\hp\hp\hm\hm\varphi\bar\varphi\>_{(e^{\dash14\phi}D^8\!R^4)_{\rm avg}}\\
    &=-2[12]^4\<34\>^4
     \Big[ \tfrac{3}{4}\sum_{i<j} s_{ij}^4 +\tfrac{1}{16} \Big(\sum_{i<j} s_{ij}^2\Big)^2\, \Big]\,.
 \end{split}
\end{equation}
However, we cannot conclude from this result that the operator $D^8R^4$ violates $E_{7(7)}$: contrary to the lower-loop cases we have studied, $D^8R^4$ is not unique. In fact, as we show later in this section, there is an infinite tower of supersymmetric  operators of mass dimension 16. It is relevant for the 6-point matrix elements that there are two independent supersymmetrizations of $D^4R^6$. To any non-linear supersymmetrization of $D^8R^4$ we can add an arbitrary linear combination of these $6$-point operators and obtain another valid supersymmetrization of $D^8R^4$. The $SU(8)$-averaged string amplitude picks out one particular such linear combination whose soft-limits~(\ref{softstringyD8R4}) happen to be non-vanishing.

We construct the matrix elements of $D^4R^6$ explicitly with Gr\"obner basis techniques and find that they have non-vanishing SSL's; specifically we find that the SSL's of the 6-point matrix elements of the operators $D^4R^6$ span the 2-parameter space
\begin{equation}\label{softD4R6}
 \begin{split}
    &\lim_{p_6\to 0}\<\hp\hp\hm\hm\varphi\bar\varphi\>_{D^4\!R^6}
    \\[-.75ex]&~~~~
    =[12]^4\<34\>^4
     \Big[ c_1\sum_{i<j} s_{ij}^4 +c_2 \Big(\sum_{i<j} s_{ij}^2\Big)^2 \,\Big]\,.
 \end{split}
\end{equation}
It follows from (\ref{softstringyD8R4}) and (\ref{softD4R6}) that we can choose a
 suitable linear combination of the two $D^4R^6$ operators to make the SSL of the 6-point matrix elements of the resulting non-linear supersymmetrization of $D^8R^4$ vanish: thus \emph{there exists a supersymmetrization of $D^8R^4$ that satisfies}
\begin{equation}\label{softD4R6b}
 \begin{split}
    &\lim_{p_6\to 0}\<\hp\hp\hm\hm\varphi\bar\varphi\>_{D^8R^4}=0\,.
 \end{split}
\end{equation}

Since this particular $D^8R^4$ satisfies the single-soft scalar theorems up to $6$ points, it is important to also analyze the \emph{double-soft} limit constraints of~\cite{Nima} that probe the structure of the coset $E_{7(7)}/SU(8)$. We numerically verified that various non-trivial double-soft limits \cite{BD} of the $6$-point matrix elements of $D^8R^4$ behave precisely as required for $E_{7(7)}$-invariance.
Therefore  the matrix elements of  $D^8R^4$ are compatible with continuous $E_{7(7)}$ up to 6 points.

We would like to alert the reader to an alternative construction of the full $n$-point superamplitudes for the
matrix elements of $n$-point $7$-loop counterterms. Once the counting of the operator's multiplicity has been established by other means (as described above), it is easy to write down a corresponding set of superamplitudes.
For the two 6-point superamplitudes of $D^4R^6$, for example,
one can choose the basis
\begin{equation}
\begin{split}\label{D4R6super}
    {\cal A}_{D^4\!R^6}&=\delta(\tilde{Q})\delta(Q)\bigl[(\varphi_1,\varphi_2)(\varphi_3,\varphi_4)\,(\varphi_5,\varphi_6)+\text{perms}\bigr]\,,\\
    {\cal A}_{D^4\!R^6{}'}&=\delta(\tilde{Q})\delta(Q)\bigl[(\varphi_1,\,\varphi_2,\,\varphi_3,\,\varphi_4,\,\varphi_5,\,\varphi_6)+\text{perms}\bigr]\,.
\end{split}
\end{equation}
Here $Q$, $\tilde{Q}$ are the usual supercharges that act on the Grassmann $\eta$-variables of the superamplitude as differentiation and multiplication, respectively, and thus
\begin{equation}
    \delta(Q)=\prod_{a=1}^8\sum_{i<j}[ij] \frac{\partial^2}{\partial \eta_{ia}\partial \eta_{ja}}\,,\quad
    \delta(\tilde{Q})=\prod_{a=1}^8\sum_{i<j}\<ij\>\eta_{ia}\eta_{ja}\,.
\end{equation}
The sums in~(\ref{D4R6super}) run over all inequivalent permutations of the external state labels $i$ of the $\varphi_i$, and the $\varphi$-products are defined as
\begin{eqnarray}
\label{prods}
    &&(\varphi_i,\varphi_j)\equiv\textstyle{\prod_{t\!=1}^4}\eta_{ia_t}\eta_{jb_t}\times\epsilon^{a_1a_2a_3 a_4 b_1b_2b_3 b_4}\,,\\[.2ex]
 	\nonumber
    &&(\varphi_i,\varphi_j,\varphi_k,\varphi_l,\varphi_m,\varphi_n)\equiv\textstyle{\!\prod_{t\!=1}^4}
    \eta_{ia_t}\eta_{jb_t}\eta_{kc_t}\eta_{ld_t}\eta_{me_t}\eta_{nf_t}\\[.5ex]
 	\nonumber
    &&\qquad~\times\epsilon^{a_1a_2b_1b_2b_3 b_4c_1c_2}\epsilon^{c_3c_4d_1d_2d_3 d_4e_1e_2}
    \epsilon^{e_3e_4f_1f_2f_3 f_4a_3a_4}\,.
\end{eqnarray}
Of course, the choice of contractions is not unique, and only through the previously established multiplicity count do we know that it is sufficient to consider the two contractions given in~(\ref{D4R6super}).
A similar construction can be carried out for the three $8$-point N$^2$MHV superamplitudes of $R^8$. Again, one can immediately propose three superamplitudes that span the space of $R^8$ counterterms, for example by considering three order-$8$ contractions involving  the $\varphi$-products~(\ref{prods}) and their $8$-scalar generalizations.

\smallsection{The infinite tower of 7-loop counterterms}
 We now examine the multiplicity of potential $7$-loop $n$-point counterterms \cite{EFK}
 \begin{equation}
    D^8R^4,~D^{4}R^6,~R^8,~\varphi^2R^8,~ \varphi^4 R^8,\ldots~\,.
 \end{equation}
  $7$-loop operators correspond to long multiplets, and are thus  supersymmetric descendants of local operators composed from only scalars with no derivatives. This follows from setting $L=7$ in (\ref{npq}) and (\ref{kL}).  $SU(8)$-singlet operators $C$ only exist for even $n$ at $L=7$, and we write them schematically as
 \begin{equation}
    C\simeq Q^{16}\tilde Q^{16} \varphi^{2q}.
 \end{equation}
The lowest weight $\varphi^{2q}$ must be in an $SU(8)$-singlet combination,
and every such singlet gives rise to a long supermultiplet.
Hence there is one 7-loop $n$-point counterterm for each singlet
in the decomposition of the symmetric tensor product of $n\!=\!2q$ $\mathbf{70}$'s.
With increasing
 $q$
there is a
(swiftly) increasing number of singlets, as illustrated by the  explicit multiplicities  up to $n=16$ in Table~\ref{ourfigure}.
Consequently, there is an infinite `tower' of independent 7-loop operators that are potential counterterms. Operators corresponding to their SSL are also listed in Table~\ref{ourfigure}. Their construction is similar, and their multiplicities is the number of $\mathbf{70}$'s in a product of  $(2q\!-\!1)$  $\mathbf{70}$'s.

\smallsection{\texorpdfstring{$E_{7(7)}$}{E7(7)} violation of higher-point 7-loop operators}

Consider the leading $n$-point matrix elements of a local counterterm $C$. If non-vanishing, the SSL produces a local $(n-1)$-matrix element, which can be generated by an $(n-1)$-point local operator $dC$ in the ${\bf 70}$. Locality of $C$ ensures that the SSL operation $C \to dC$ commutes with the SUSY generators $Q$ and $\tilde{Q}$. For a long-multiplet ($L\geq7$) counterterm $C=Q^{16}\tilde Q^{16} C_0$ we can therefore write
\vspace{-2mm}
\begin{equation}
dC=Q^{16}\tilde Q^{16} dC_0.
\end{equation}
$E_{7(7)}$ requires that the SSL vanishes. Now there are two ways to obtain $dC=0$:
either the single-soft limit of $C_0$ vanishes ($dC_0=0$), or $dC_0$ is annihilated by $Q^{16}\tilde Q^{16}$.
At the seven-loop level, $C_0=\varphi^{2q}$ consists of only scalars with no derivatives, and
consequently $dC_0\neq0$.
$C_0$ is also not annihilated by $Q^{16}\tilde Q^{16}$
because $dC_0=\varphi^{2q-1}$ in a $\mathbf{70}$
satisfies a shortening condition only for $n=2q\leq 4$
\ecite{\shorts}.
Therefore, all 7-loop linearized counterterms with $n>4$ have
non-vanishing single-soft scalar limits and thus violate $E_{7(7)}$. (This was also observed in~\cite{Kallosh:2010kk}; see~\cite{Bianchi:2009wj} for discussion of non-perturbative aspects.)
These operators may, however, play an important role as dependent terms in
the non-linear completion of the $D^8R^4$ operator, as we  demonstrated above at the $6$-point level.

\vspace{-0.5mm}
\section{SSL structure: 7-, 8- and 9-loops}
\vspace{-0.5mm}
We now show that all our findings on $E_{7(7)}$-(in)com\-pa\-ti\-bi\-li\-ty of operators have a natural explanation in terms of the multiplicities of SSL operators in the ${\bf 70}$
that is displayed in Table~\ref{ourfigure}.\footnote{Throughout this section we are only concerned with the lowest-point  non-vanishing SSL's of an operator. If an operator has non-vanishing SSL's at $n$-point, its higher-point matrix elements will generically have {\em non-local} SSL's, which are not classified by our analysis.}
Let us first revisit the case of $D^4R^4$ and $D^6R^4$. At the $5$- and $6$-loop level, there are no potential $3$-point or $4$-point SSL operators available. Therefore the  matrix elements of $D^4R^4$ and $D^6R^4$ must have vanishing soft limits at $4$- and $5$-points, and this is indeed the case. There exists, however, one potential $5$-point SSL operator at $L=5$ and $L=6$. Generically, one expects the soft-limits of the $6$-point matrix elements of $D^4R^4$ and $D^6R^4$ to be proportional to this operator with some non-vanishing coefficient. This is precisely what happens.

At $7$ loops with $n>4$ points, the number $n_{S}$ of SSL operators $D^{16}\varphi^{n-1}$ is always at least as large as the number $n_C$ of potential counterterms $D^{16}\varphi^{n}$. Generically, one therefore expects the soft limits of the potential counterterms to  span an $n_C$-dimensional subspace in the $n_S$-dimensional space of SSL operators. It would follow that all potential counterterms with $n>4$ violate $E_{7(7)}$.  Indeed, this is what we explicitly proved above for the $7$-loop case.

By the same logic, it is not at all surprising that  there is a non-linear supersymmetrization of $D^8R^4$ that preserves $E_{7(7)}$ at the $6$-point level. The number of $5$-point SSL operators precisely matches the number of $D^4R^6$ operators. Therefore, the $6$-point soft limit of $D^8R^4$ can be made to vanish after adding an appropriate combination of the two $D^4R^6$ operators, just as we found above. For the $8$-point soft limits of $D^8R^4$, however, there are 4 SSL operators available; more than the $3$ potential $8$-point counterterms $R^8$. If the  $8$-point soft limits of $D^8R^4$ take a generic value in the $4$-dimensional space of SSL operators, no linear combination of $R^8$ operators can be chosen to give an $E_{7(7)}$-preserving supersymmetrization of $D^8R^4$; a remarkable cancellation is thus required for $D^8R^4$ to be compatible with $E_{7(7)}$.

As Table~\ref{ourfigure} illustrates, $E_{7(7)}$ becomes more and more constraining as we increase the number of points and loops. For example, the $14$-point soft limits of $D^{10}R^4$ have to lie in a specific $153$-dimensional subspace of the $1033$-dimensional space of SSL operators in order for $D^{10}R^4$ to satisfy $E_{7(7)}$ after an appropriate addition of independent $14$-point operators. It follows that $E_{7(7)}$ is a very constraining symmetry even for $L=7$ and beyond. Although there is an infinite tower of independent counterterms at each of loop $L\geq7$, we cannot expect any of these operators to preserve $E_{7(7)}$ `accidentally'. There may, however, be a very good reason for the cancellations of terms that is needed for $E_{7(7)}$-invariant operators to exist for $L\geq 7$; namely, when there is a construction of a manifestly $E_{7(7)}$-invariant supersymmetric operator~\ecite{L7counterterms,\LBCounterterms}. At the $7$-loop level, for example, we can only expect the $8$-point single-soft limits of $D^8R^4$ to vanish after an appropriate addition of $R^8$, if the manifestly $E_{7(7)}$-invariant superspace integral that was proposed as a candidate counterterm in~\cite{L7counterterms} is indeed non-vanishing.

One new feature that emerges at $L=8,9$ is the existence of $n>4$ operators
that have vanishing soft-limits at the linearized level.  This holds for the MHV operators $D^8R^5$, $D^{10}R^5$ and $2\times D^8R^6$ as well as for at least $7$ of the $12\times D^8R^6$ NMHV operators. The latter follows 
 from the multiplicity $5$ of $5$-point SSL operators at 9 loops.  $E_{7(7)}$-invariance beyond the linearized level, however, is a highly non-trivial constraint on all of these operators.

 \section*{Acknowledgements}
We thank M.~Headrick for suggesting the use of the Gr\"obner basis for the study of counterterms in supergravity.
We thank T.~McLoughlin,  S.~V~Sam and P.~Vanhove for valuable discussions.

The research of DZF is supported by NSF grant PHY-0600465 and by
the US Department of Energy through cooperative research agreement DE-FG-0205FR41360.
HE is supported by NSF CAREER Grant PHY-0953232, and in part by the US Department of Energy under DOE grants DE-FG02-95ER40899 (Michigan) and DE-FG02-90ER40542 (IAS).
The research of MK is supported by the NSF grant PHY-0756966.



\raggedright

\end{document}